\documentclass[twocolumn,showpacs,preprintnumbers,amsmath,amssymb]{revtex4}
\usepackage{amsmath}

\usepackage{graphicx}
\usepackage{dcolumn}
\usepackage{bm}

\begin{document}

\title{Measuring Significance of Community Structure in Complex Networks}

\author{Yanqing Hu\footnote{yanqing.hu.sc@gmail.com}, Yuchao Nie, Hua Yang, Jie Cheng, Ying Fan and Zengru Di\footnote{zdi@bnu.edu.com}}

\affiliation{Department of Systems Science, School of Management,
Center for Complexity
 Research, Beijing Normal University, Beijing 100875, China}

\begin{abstract}

Many complex systems can be represented as networks and separating a
network into communities could simplify the functional analysis
considerably. Recently, many approaches have been proposed for
finding communities, but none of them can evaluate the communities
found are significant or trivial definitely. In this paper, we
propose an index to evaluate the significance of communities in
networks. The index is based on comparing the similarity between the
original community structure in network and the community structure
of the network after perturbed, and is defined by integrating all
the similarities. Many artificial networks and real-world networks
are tested. The results show that the index is independent from the
size of network and the number of communities. Moreover, we find the
clear communities always exist in social networks, but don't find
significative communities in proteins interaction networks and
metabolic networks.
\end{abstract}

\keywords{Complex Network, Community Structure, Robustness}

\pacs{89.75.Hc, 87.23.Ge, 89.20.Hh, 05.10.-a}

\maketitle

The study of the community structure of networks has become a very
important part of researches of complex networks. Nodes belonging to
a tight-knit community are more likely to have particular properties
in common. In social relationship network, communities usually
represent different friend subgroups. In the world wide web,
community analysis has uncovered thematic clusters. In biochemical
or neural networks, different communities may represent different
functional groups, and separating the network into such groups could
simplify the functional analysis considerably. As a result, the
problem of identification of communities has been the focus of many
recent efforts. So two questions are proposed, the first is, how to
detected communities in the networks? In recent studies, plenty of
algorithms are proposed \cite{1,Newman reviwe,social
network2,3,4,6,9,10,12,14,17,18,GN,Newman mixture algorithm,EO} (see
\cite{4} as a review). The second question is coming hand in hand
with the first question: how to evaluate the communities detected?
We believe that there exist clear communities in some networks while
no clear communities in the other networks. But almost all
algorithms could find the ``community structure" in networks in
their ways, without thinking about whether the community structure
actually exists or not. Even many algorithms can also find the
community in random networks, in which are considered having no
community. For the existence of such a situation, the discussion on
the ``significative communities" is needed. As a network is given,
it is meaningless to detect the community when the community
structure is not significative at all.

Scientists try to propose a universal index to evaluate the
partitions. And the modularity $Q$ \cite{Newman evaluating
algorithm} was presented as an index of community structure and by
now it has been widely accepted \cite{4,14,17,EO} as a measure for
the community structure. Modularity $Q$ was presented as a index of
community structure by Newman and Grive, which was introduced as
$Q=\sum_{r}{(e_{rr}-a_{r}^{2})}\label{Q}$, where $e_{rr}$ are the
fraction of links that connect two nodes inside the community $r$,
$a_{r}$ the fraction of links that have on or both vertices in side
the community $r$, and sum extends to all communities $r$ in a given
network. The larger the value of $Q$ is, the more clearly a
partition into communities is. Hence, the value of the modularity
can be used as a significative index for communities. Unfortunately,
despite the obvious advantages of modularity, it has its own
problem. It is true that networks with strong community structure
have high modularity but not all networks with high modularity have
strong community structure \cite{high Q low m}. Here, we just say
$Q$ value is not a very good index to evaluate the significance of
community structure, but do no mean that maximizing modularity $Q$
cannot detect community structure. Many empirical and numerical
results represent maximizing modularity $Q$ is a good method for
detecting communities \cite{4,14,EO}. Therefore in the following
analysis, we still use maximizing modularity $Q$ to detect community
structure.

Recently, Karrer, Levina, and Newman have suggested a method to
perturb the networks. They have shown some phenomena about the
robustness of community structure in networks \cite{robustness}.
Intuitively, if a network has distinct communities, the community
structure should be robust under perturbation. Thus in this paper,
we develop a perturbation method and propose an index to measure the
significance of communities based on the perturbation to the
network, and try to solve the second question mentioned above. In
our method, we strengthen the perturbation to the network from just
small amount of edges rewired to all edges rewired. Then we can get
the results of perturbations (the similarity of community structure
between original and perturbed networks) for each case. Finally we
get our index by integrating all the similarities of perturbations.
Using our index, we can evaluate whether the network has a
``significative communities". The method is described in detail in
the following section. Naturally, we apply the method to many kinds
of networks, and find some interesting conclusions. We argue that
social networks usually have distinct and significative community
structure, while metabolic networks also have community structure
but not so clear. However, some protein interaction networks we
tested have no significant communities.

\section{Method}
There are three steps to get our index for a given network. First,
we detect the communities in the original network without any
perturbation. Second, we will perturb the network, using the way of
perturbing the edges in network by an arbitrary amount. Then we can
detect the new communities after perturbation. Besides, we calculate
the similarity between the two partitions (the communities of
original network and perturbed network). Third, we increase the
proportion of edges perturbed little by little until all edges are
perturbed, repeat the process of second step, and compare the new
communities with original ones with perturbation strengthened.
Hence, we can get a series of proportion of perturbation as well as
the corresponding similarity values. At last, we sum up all the
products of similarity values and the corresponding increased
proportion of perturbation. If we just increase the proportion
little enough, the process just like the calculation of integration.

When we perturb the edges in the network, there are various methods
to achieve. In this paper, we adopt absolutely random perturbation
to the network. Consulting to the method of network perturbation
introduced by Newman \cite{robustness}, we makes sure the total
number of edges is unchanged, which make the comparison of the
partitions straightforward. Specifically, we go through each edge in
original network and with probability $p$ we remove it, then we add
the same amount of edges randomly between any two nodes, which have
no connection after perturbation. In this way, if $p$=0, no edge is
moved and the network is all the same with original. If $p$=1, all
edges are moved and the process generates a random graph, which has
no correlation with original. And for values of $p$ between 0 and 1
the perturbation generates networks in which some of the edges
retain their original positions while the others are moved to new
positions. Therefore, we adopt a sequence perturbation to the
network. We do not only perturb networks by little, but also
strengthen the proportion of perturbation until all the edges move
their positions. Further, we do not care if the expected average
degrees of every node is the same as before, which is different from
Karrer and Newman et al\cite{robustness}. We argue that the
absolutely random perturbation to the network is more reasonable,
simple and efficient.

After detecting the community structure in the networks perturbed,
the question becomes how to compare the similarity between the
communities perturbed and the original. We think that a more
discriminatory measure is the normalized mutual information index,
which is based on information theory, as described in Ref
\cite{Systematic}. They defines a confusion matrix $N$, where the
rows correspond to the ``real" communities in networks without
perturbation, and the columns correspond to the ``found"
communities. The element of $N$, $N_{ij}$ is the number of nodes in
the real community $i$ that appear in the found community $j$.
Therefore a measure of similarity between the partitions $A$ and $B$
is
\begin{equation}I(A,B)=\frac{-2\sum_{i=1}^{c_A}\sum_{j=1}^{c_B}N_{ij}log(\frac{{N_{ij}}{N}}{{N_{i.}}{N_{.j}})}}{\sum_{i=1}^{c_A}N_{i.}log(\frac{N_{i.}}{N})+\sum_{j=1}^{c_B}N_{.j}log(\frac{N_{.j}}{N})}\end{equation}
As the discrepancy of partitions increases, the value of $I(A,B)$
decreases from 1. In this paper, we compare the ``communities
without perturbation" $A$ and the ``communities after perturbation"
$A(p)$, what is different, we make a little change on the similarity
index. We found that the $I(A,A(p))$ has been not only decided by
the discrepancy of the communities, but also influenced by the size
of networks and the number of community in $A$ and $A(p)$. In order
to eliminate the influence of the size, we consider the improved
measure below:
\begin{equation}S(A,A(p))=I(A,A(p))-I(A_{rand},A_{rand}(p))\end{equation}
where, $A_{rand}$ or $A_{rand}(p)$ has same number of communities
with $A$ or $A(p)$, moreover each community in $A_{rand}$ or
$A_{rand}(p)$ has the same number of nodes with the corresponding
community in $A$ or $A(p)$ respectively. But different from $A$,
$A(p)$ that are correlated with the original network, the nodes in
$A_{rand}$ and $A_{rand}(p)$ are randomly selected form the whole
set of nodes. In this way, we can get a series of values of
$S(A,A(p))$ by strengthening the proportion of perturbations from 0
to 1 little by little. We adopt $0.02$ as the increased proportion
of perturbation for each time in this paper. Generally, a higher
proportion of perturbation corresponding to a lower value of
$S(A,A(p))$. Hence,we can get our measure as following:
\begin{equation}R=\int_{0}^{1}S(p)dp\end{equation}
where $p$ is the proportion of perturbation, and $S(p)$ is the
similarity value between original community structrue and the
community sturcture when the proportion of perturbation is $p$. If a
network has distinct community structure, the value of our measure
$R$ is inclined to high. On the contrary, the network holding fuzzy
community structure displays low value. For a random network $R$
will approach to 0 theoretically. The value of the similarity is a
function of the parameter $p$ that measures the amount of
perturbation. The similarity value starts at 1 when $p=0$, as we
would expect for an unperturbed network. Then the similarity value
drops off and approaches its minimum value while $p=1$, while the
network at present is an absolute random network.

\begin{figure}
\includegraphics[width=8cm]{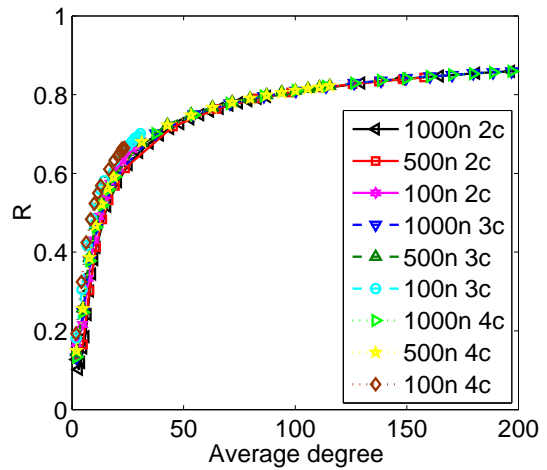}
\caption{The relationship among the value of $R$ index, network size
average degree and community number. In the plot `$x$n $y$c' denotes
$x$ nodes and $y$ pre-determined communities with same size. Every
pre-determined communities (sub-networks) are generated by the same
way. They are ER networks and disconnected with each other. From the
plot we can see that the value of $R$ increase with the increasing
of average degree and is almost independent from the size of network
and number of communities.}\label{234_comm}
\end{figure}

\begin{figure}
\includegraphics[width=8cm]{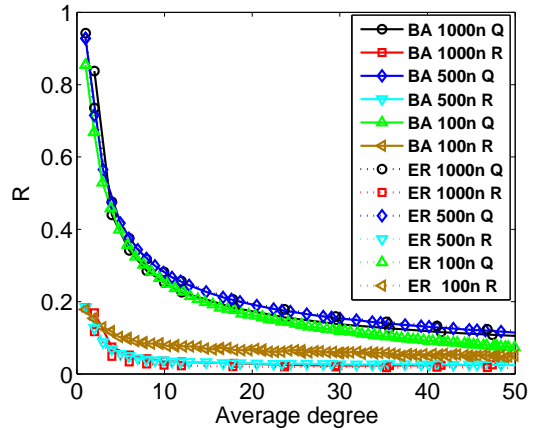}
\caption{Comparing modularity $Q$ and index $R$ in BA and ER
networks in which there exists no community structure. From the plot
we can see that, $Q$ is very large when the average degree is about
1, while, value of $R$ is near 0.2. That is to say, when average
degree is near 1, $Q$ index presents very strong community
structure, and $R$ shows fuzzy community structure (we obtain that
there exits fuzzy community structure when $R=0.2$ form the
numerical results in artificial networks (see Fig.
\ref{artificial_network})). But when the average degree increases,
$Q$ drops more slowly than $R$. When the average degree is larger or
equal to 2, $R$ is very low and achieves stable state soon, which
indicates $R$ index perform well in both BA and ER networks.
}\label{QI}
\end{figure}

Dose the measurement is independent with the size of network, and
what will happen when changeing the number of communities with same
size and number of edges? Moreover, can the measurement work well in
some networks that $Q$ index fails to measure \cite{high Q low m}?
In order to give answers to the above questions, firstly we apply
the measurement $R$ in same size networks with same number of
communities, and each community with same number of nodes. There are
no edge between different sub-networks and each of them is ER
network. That means the communities are distinct. Numerical
experiments present that, the value of index $R$ is roughly
independent with the size of network and number of communities. When
the average degree increases, the value of $R$ will increase
correspondingly (as shown in Fig. \ref{234_comm}). Secondly, we
compare $Q$ index and $R$ index in ER networks. It is known that $Q$
index cannot measure ER and BA networks \cite{high Q low m}. For the
BA and ER networks with lower average degree, the modularity $Q$
could be very high. So we compare $R$ index with $Q$ index in
different BA and ER networks with different size and average degree.
The results tell us that $R$ index has the same behaviors in BA and
ER networks. When the average degree is large or equal to 2, $R$
index will be lower than 0.1, and soon be stable. When the average
degree is 1, the $R$ is less than 0.2. From the following
applications on artificial networks (as shown in Fig.
\ref{artificial_network}), we known that $R<0.1$ is a low value. It
indicates there are no community structure in the network. But
$R=0.2$ is not very low. It presents there exists fuzzy communities
in the network. Hence, our index performs well but it is also not
suitable for some networks where average degree is less than 1.
Fortunately, there are few real-world networks with average degree
less than 1. By and large, our index is more efficient than $Q$
index in BA and ER networks (as shown in Fig. \ref{QI}). Moreover,
from the numerical experiments we find that for a very large size
network which contains two equal clique-complete community structure
network, the value $R$ can be larger than 0.9, and the value of $R$
can lower than 0.03 for large size random networks with proper
average degree. Thus, we can conclude that $R\in (0,1)$ roughly.

\section{Result}
In order to test the validity of our index. Firstly, we apply it on
computer-generated random networks with a well-known predetermined
community structure. Each network has $n=128$ nodes divided into $4$
communities of $32$ nodes each. Edges between two nodes are
introduced with different probabilities depending on whether the two
nodes belong to the same community or not: every node has $\langle
k_{in}\rangle$ links on average to its fellows in the same
community, and $\langle k_{out}\rangle$ links to the other
communities, keeping $\langle k_{in}\rangle+\langle
k_{out}\rangle=16$. As is known to all, the communities become more
and more diffuse and harder to identify when $k_{out}$ increase,
hence the significance of the communities found by algorithm also
tends to weakness and $R$ index will decrease. In order to validate
the expectation that $R$ index will become lower as the $k_{out}$
decreases, we calculate the value of $R$ in the case that $k_{out}$
ranges from $0$ to $12$. The method we use to detect community
structure in this paper is the combination of Newman's spectral
algorithm and extremal optimization algorithm \cite{14,EO}. We use
spectral algorithm to detect the initial community structure, and
extremal optimization algorithm to improve the community partition.
We also use an other algorithms \cite{signal} to detect the
community structure, we find the influence of different algorithms
on our index is neglectable.

\begin{figure}
\center
\includegraphics[width=8cm]{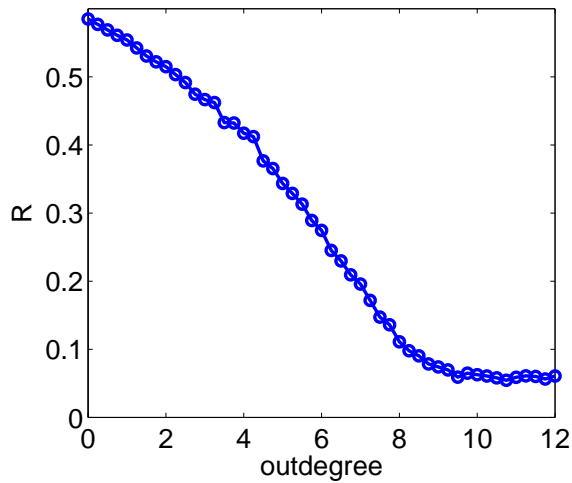}
\caption{ The $x$-axis is $k_{out}$ (the proportion of connections
between communities),while the $y$-axis represents the value of
measurement as described in this paper. The percentage we increase
the proportion of perturbation is $0.02$ for each time. Each value
corresponding to the $k_{out}$ is the average value of $20$
numerical experiments where each time we generate a new independent
network. The value of $R$ is $0.58$ when $k_{out}$ is $0$ but $R$ is
$0.05$ when $k_{out}$ is $11$. When $k_{out}$ is about 11, the
network is random in which there is no community structure
theoretically.}\label{artificial_network}
\end{figure}

The result is shown in Fig.\ref{artificial_network}. As our
anticipation, the value of index varies from $0.58$ to $0.05$ as
$k_{out}$ varies from $0$ to $12$, which means that our index have
good ability to mark the significative of communities in
compute-generated network. For larger values of $k_{out}$, the value
of the index is lower, indicating that the community structure is
not more significative than that of a random graph. The index
decreases as a function of $k_{out}$, indicating that the community
structure discovered by the algorithm is relatively significative
when $k_{out}$ is relatively low (or $k_{in}$ is relatively high).

\begin{figure}
\center
\includegraphics[width=8cm]{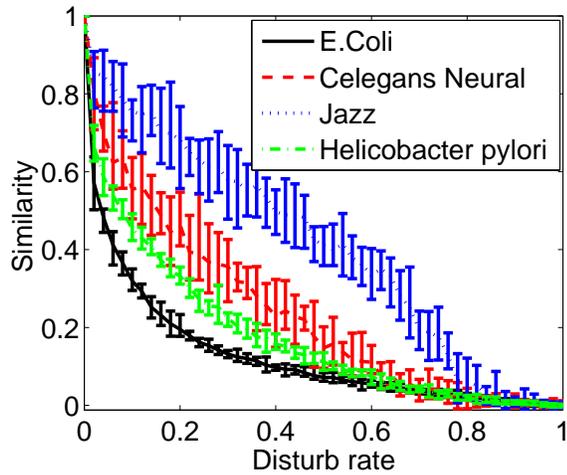}
\caption{The $x$-axis represent the average number of perturbation,
while the y-axis is the similarity value $S(p)$. For each network,
we earmark three value here: maximum, minimum and mean of 20 times
simulation. The network in turn are proteins interaction network (E.
coli), neural network (C. elegans), social network (Jazz) and
metabolic network (Helicobacter Pylori) from the bottom up. The
increase of he proportion of perturbation every time is
0.02}\label{New_4_networks}
\end{figure}

Of course, we also apply it on many real networks
\cite{ZK_d,Dolphin,Ploblog,C.neural,Jazz,C.met,webset1,webset2,webset3}.
A good index shouldn't be available to computer-generated networks
only, but also has good behavior in real networks. It is necessary
to proof-test our index on all kinds of real networks. People
usually classify the real networks into three sorts: social networks
(such as scientist collaborations and friendships), biological
networks (such as proteins interaction networks and metabolic
networks) and technological networks (such as Internet and the WWW).
Distinct communities within networks have been observed in different
kinds of networks, most notably in social networks while  fuzzy in
biological networks often. We apply the index into many different
networks, and obtain relatively high value of our index in social
networks. Therefore, we validate the availability of our index. You
can get more detail form Fig.\ref{New_4_networks} and Tab.\ref{Table
real networks}. Fig.\ref{New_4_networks} shows the curves of $S(p)$,
using 4 networks as an example. Here we average the results of the
20 times simulation in the figure, in which we earmark the maximum,
minimum, and the mean value of the 20 times simulation. As is shown,
the similarity measure of Jazz network decrease slowly while the
similarity of the other three networks decrease rapidly. The figure
argues that the communities in Jazz network are more robust than
other three. It means that the structure of the Jazz network is
hardly changed under perturbation. Thus the community structure in
Jazz network is distinct and significative. Tab.\ref{Table real
networks} shows all the networks we apply the index on. From the
table, we find different kinds of networks have different index
values, which indicate the significance of the communities in
different networks varies. First, we analyze several social
networks, including Zachary karate club network, dolphin network,
collage football network, Jazz network, scientists collaboration
network and so on. We get relatively high value of our index among
these networks, and most of these networks have the index value over
0.27, which shows the existence of strong community structure in
these networks, and the community structure found in these networks
are clear. However, the Santa Fe scientists collaboration network
has an index value 0.14, which is low. As is known, the Santa Fe
Institute is different from many other Institutes. Renowned
scientists and researchers that come to Santa Fe Institute are from
universities, government agencies, research institutes, and private
industry. Therefore the relationship between the members is not as
tight as other collaboration networks. All the social networks in
Tab.\ref{Table real networks} are networks of friendship
(collaboration could be viewed as a form of friendship). Just as
said ``Birds of a feather flock together", it is easy to understand
why the social networks always centralize as some distinct groups.

What's more, we also analyze some biological networks such as
proteins interaction networks E.coli, Yeast and H.Sapiens, and many
metabolic networks. We find in proteins interaction networks the $R$
index value are low (the average degree of H.Sapiens is less than 2,
so its $R$ index is little high). In metabolic networks, we
calculate 43 metabolic networks, all the index value $R$ are medium
about 0.19. For the average nodes' number 1488 and average edges'
number 3460, the average index value is 0.19. Therefore we conclude
that in some proteins interaction networks (such as E.coli and
Yeast) and the metabolic networks which are listed in the following
table, there are no clear communities. It may be unnecessary to
detect and analyze the communities in these networks.

\begin{table}
\caption{The integral measure of some real networks. The table shows
the names of different real networks and the corresponding index
values. The column of size denotes the number of nodes and
edges}\label{Table real networks}

\begin{tabular} [t]{|c|c|c|c|c|}
\hline network&size&$R$&type\\\hline
E.coli&1442, 5873&0.14&protein\\
Yeast&1870, 4480&0.14&\\
H.Sapiens&693, 982&0.21&\\\hline
Celegans metabolic&453, 4596&0.19&metabolic\\
Aquifex aeolicus&1485, 3400&0.19&\\
Helicobacter pylori&1363, 3151&0.19&\\
Yersinia pestis&1950, 4505&0.18&\\
43 metabolic networks&1488, 3460&0.19&\\\hline Celegans neural&297,
2359&0.24&neural\\\hline
Santa Fe scientists&260, 2692&0.14&social\\
Zachary karate&34, 78&0.27&\\
Dolphin&62, 159&0.27&\\
College
football&115, 613&0.38&\\
Jazz&198, 5484&0.42&\\
Political blogs&1224, 19090&0.29&\\
Political books&105, 441&0.34&\\\hline
\end{tabular}

\end{table}

\section{Conclusion and discussion}
In this paper an index is presented which can measure the
significance of communities detected. The index is based on
comparing the similarity between the original community structure
and the community structure after perturbed in the network. Then the
index value is the integration of all the similarities. We apply the
index to many artificial and real world networks, such as social
networks, neural network, proteins interactions networks and
metabolic networks. The results show that our index is independent
form the network size and community number. Moreover we find the
different kinds of networks have different characteristics, social
networks usually have significant communities, while communities are
comparatively fuzzy in biological networks, especially in some
protein-interaction networks.

\section*{Acknowledgement}
This work is partially supported by 985 Project and NSFC under the
grant No. 70771011, and No. 60534080.

\end{document}